%% file: main.tex
\documentclass[letterpaper,twocolumn,10pt]{article}
\usepackage{usenix}

\usepackage{tikz}
\usepackage{amsmath,amssymb,amsfonts}

\usepackage{makecell}
\usepackage{booktabs} 

\usepackage{filecontents}
\usepackage{cite}

\usepackage{algorithmic}
\usepackage{graphicx}
\usepackage{textcomp}
\usepackage{xcolor}
\usepackage{subfig}
\usepackage{multirow}
\usepackage[ruled,vlined,linesnumbered]{algorithm2e}

\SetCommentSty{mycommfont}

\newcommand*{\Mname}{\textit{WLB-LLM}}

\begin{document}

\date{}

\title{\Large \bf WLB-LLM: Workload-Balanced 4D Parallelism for \\ Large Language Model Training}

\author{
{\rm Zheng Wang$^{1,2}$, Anna Cai$^2$, Xinfeng Xie$^2$, Zaifeng Pan$^{1}$, Yue Guan$^{1}$, Weiwei Chu$^2$, Jie Wang$^2$,} \\ {\rm Shikai Li$^2$, Jianyu Huang$^2$, Chris Cai$^2$, Yuchen Hao$^2$, Yufei Ding$^{1,2}$}\\
University of California, San Diego$^1$ \\
Meta$^2$
}

\maketitle

\begin{abstract}
In this work, we present \textit{WLB-LLM}, a \textit{\underline{W}}ork\textit{\underline{L}}oad-\textit{\underline{B}}alanced 4D Parallelism for \textit{\underline{L}}arge \textit{\underline{L}}anguage \textit{\underline{M}}odel Training. We first thoroughly analyze the workload imbalance issue in LLM training and identify two primary sources of imbalance at the pipeline parallelism and context parallelism levels.
Then, to address the imbalance issue, at the pipeline parallelism level, \textit{WLB-LLM} incorporates a workload-aware variable-length document packing method to balance the computation and communication workload across micro-batches. Additionally, at the context parallelism level, \textit{WLB-LLM} introduces a novel fine-grained per-document sharding strategy, ensuring each worker within a context parallelism group has an identical workload.
Comprehensive experiments under different model scales demonstrate that \textit{WLB-LLM} significantly mitigates the workload imbalance during 4D parallelism LLM training and achieves an average speedup of $1.23\times$ when applying \textit{WLB-LLM} in our internal LLM training framework.
\end{abstract}

\input{sections/01_Introduction}
\input{sections/02_background}

\input{sections/03_motivation}

\input{sections/04_PP_level}

\input{sections/05_CP_level}

\input{sections/06_evaluation}

\input{sections/07_related_work}

\input{sections/08_conclusion}

\bibliographystyle{plain}
\bibliography{refs}
\end{document}

%% file: sections/01_Introduction.tex
\section{Introduction}

Large language models (LLMs) have been widely adopted as the backbone of various applications, such as coding assistants~\cite{wei2023copiloting, nijkamp2022codegen}, language translation~\cite{zhang2023prompting}, and chatbots~\cite{chiang2023vicuna, OpenAI_ChatGPT}. 
The remarkable capabilities and promising potential of LLMs have sparked a competition among big tech companies to train LLMs with higher quality and greater capabilities~\cite{achiam2023gpt, team2023gemini, dubey2024llama}. 
As the scale of LLMs and length of context window continue to grow larger~\cite{kaplan2020scaling}, the training of LLMs consumes a significant amount of computing power~\cite{perrault2024artificial}. For instance, Meta reports that the training of the LLaMA3-405B model uses 16K H100 GPUs for several months~\cite{dubey2024llama}.
The tremendous computational cost of LLM training makes every improvement in end-to-end training efficiency translate to substantial savings.

\begin{figure}[t] \small
    \centering

    \subfloat[Normalized computation latency in an 8K-GPU LLM training job.]{\includegraphics[width=0.99\columnwidth]{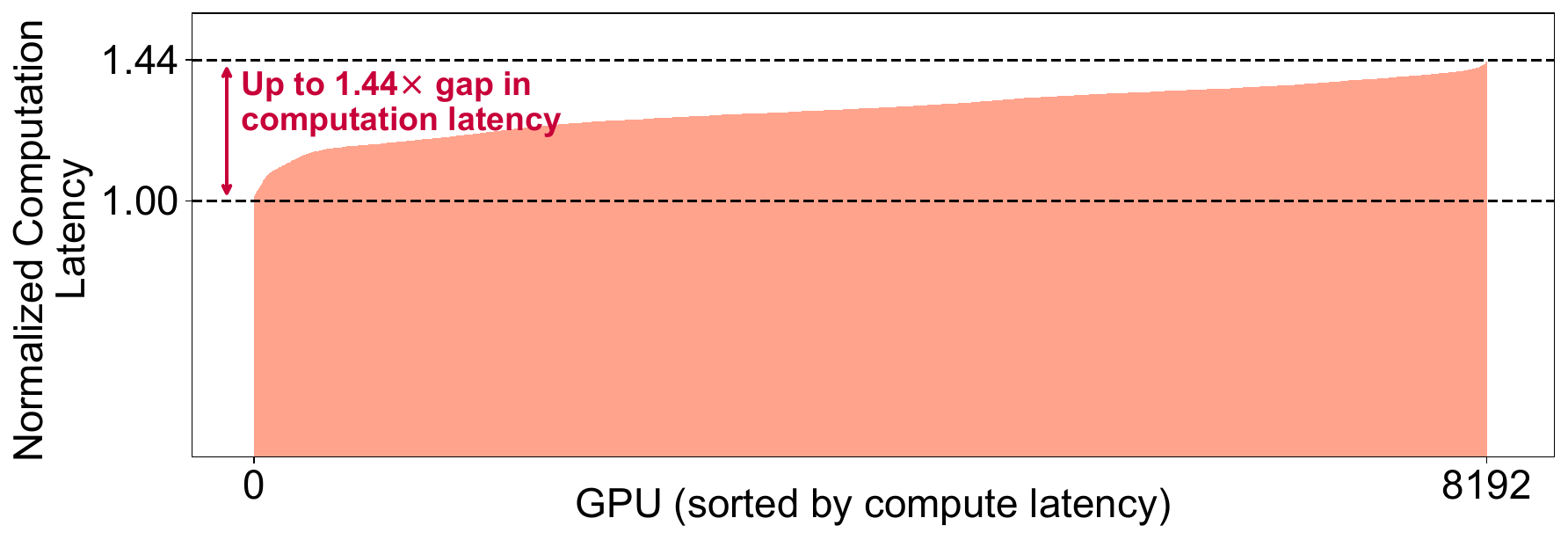}} 

    \subfloat[Reason of imbalance: input-dependent nature of attention computation and the varying input document length.]{\includegraphics[width=0.99\columnwidth]{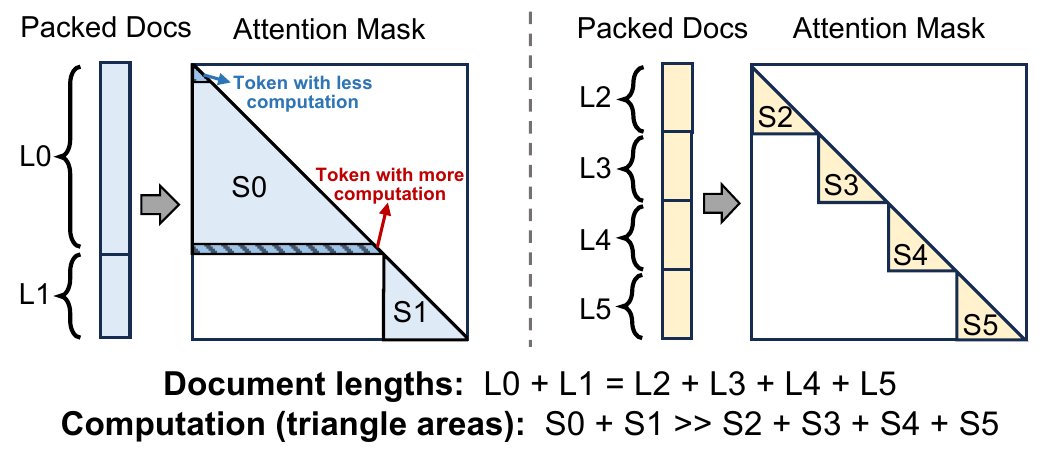}} 

    \caption{Observed workload imbalance issue in large-scale LLM training jobs and the reason of workload imbalance.}
    \label{fig: imba_motivation}
    
\end{figure}

Unfortunately, a significant portion of GPUs are underutilized during large-scale LLM training jobs due to the workload imbalance issue. To illustrate this, we present the normalized computation latency on each GPU in our internal 8K-GPU LLM training job with a context window of 128K in Figure~\ref{fig: imba_motivation} (a). 
It can be observed that the computation latency exhibits significant variance among GPUs. The slowest GPU experiences a computation latency that is $1.44\times$ longer than the others. This disparity in computing latency causes a substantial degradation in training efficiency as the synchronized nature of training requires all other GPUs to wait for the slowest GPU to finish.

The root cause of the workload imbalance lies in the input-dependent nature of attention computation and the varying input document length of the training data.
As shown in Figure~\ref{fig: imba_motivation} (b), input sequences are composed of multiple input documents. To ensure model quality in long-context pre-training, attention masks are applied to prevent self-attention computation between tokens from different documents within the same sequence~\cite{dubey2024llama}.
This approach introduces heterogeneity in per-token arithmetic intensity, as tokens at the tail positions of long documents require more attention computation. As a result, input sequences containing long documents incur significantly higher computation workloads, even with the same total sequence length.

However, existing LLM training frameworks~\cite{rajbhandari2020zero, shoeybi2019megatron, narayanan2021efficient, smith2022using, liu2024blockwise} fail to recognize the heterogeneity in per-token arithmetic intensity.
Specifically, state-of-the-art LLM training solutions employ a 4D parallelism paradigm that combines data parallelism (DP)\cite{rajbhandari2020zero}, pipeline parallelism (PP)\cite{narayanan2019pipedream}, context parallelism (CP)\cite{nvidia2023megatron}, and tensor parallelism (TP)\cite{shoeybi2019megatron}.
The input documents are packed into sequences with fixed length at the DP and PP levels, and then sharded and distributed into chunks at the CP and TP levels. 
This fixed and static training flow treats all input tokens homogeneously and assigns each GPU an equal number of tokens, inevitably resulting in workload imbalance between GPUs.
Furthermore, the trend of larger models and longer context windows exacerbates this issue, increasing the likelihood of an extremely long document appearing in the input batches, thereby delaying the entire training step.

{An intuitive approach to address the workload imbalance issue is to shuffle and repack input documents to distribute the computation workload more evenly across micro-batches. However, this method is not effective and practical for two main reasons:
\textit{Firstly}, achieving effective balance through shuffling and repacking requires a sufficiently large packing window spanning multiple global batches. This impacts the randomness of data sampling and loading, which can potentially affect model convergence during training (as discussed in Section~\ref{sec: tradeoff}).
\textit{Secondly}, shuffling and repacking only address workload imbalances across micro-batches and cannot resolve intra-document imbalances caused by sequence sharding. In 4D parallelism training, input sequences are divided into chunks and distributed across different GPUs. Document chunks that include the tail end of a document incur a higher computational workload because their tokens must attend to more preceding tokens, causing an intra-document workload imbalance across GPUs.}

To overcome the challenges mentioned above and address the severe workload imbalance issue in large-scale LLM training, we propose a flexible and input-aware document packing and sharding approach for the 4D parallelism training paradigm.
The data packing and sharding will no longer output micro-batches with a fixed number of tokens. Instead, each GPU aims to get input tokens that have an equal amount of total computation and communication workload.
Additionally, to minimize the impact on data randomness, we propose to only adjust the execution order of extremely long documents. This is based on our observation that the tokens of long documents account for only a small proportion of the total training tokens but have the most significant impact on workload imbalance.

Based on the above design insight, we build {\underline{\Mname}}, a \textit{\underline{W}}ork\textit{\underline{L}}oad-\textit{\underline{B}}alanced 4D Parallelism for \textit{\underline{L}}arge \textit{\underline{L}}anguage \textit{\underline{M}}odel Training.
We begin by thoroughly analyzing workload imbalance in LLM training under 4D parallelism (\S\ref{sec: motivation}), identifying two primary sources of imbalance across specific parallelism hierarchies: (1) Imbalance across micro-batches at the pipeline parallelism level and (2) Imbalance across document shards at the context parallelism level.
To address these imbalances, {\Mname} provides novel solutions tailored to each level.
\textbf{At the PP level}, {\Mname} introduces variable-length document packing, allowing shorter documents to be combined to form longer sequences, thereby aligning the total computation workload to that of a single long document (\S\ref{sec: PP}). 
Additionally, {\Mname} adaptively delays the execution of extremely long documents, achieving near-optimal workload balance across micro-batches while maintaining a relatively low per-token delay, thereby preserving the randomness of the data loader.
\textbf{At the CP level}, {\Mname} incorporates a novel per-document sharding strategy, ensuring each worker within a CP group has an equal workload (\S\ref{sec: CP}). 
Furthermore, we observe a tradeoff between kernel efficiency and sharding balance with per-document sharding. To maximize overall performance, we propose a heuristic algorithm that adaptively selects the most efficient sharding strategy based on the input sequence at runtime.

In summary, this paper makes the following contributions:
\begin{itemize}
    \item To the best of our knowledge, we are the first to identify, analyze, and address workload imbalance issues in large-scale LLM training with 4D parallelism.

    \item At the PP level, we design a variable-length input packing and adaptive outlier delay strategy to achieve near-optimal workload balance across micro-batches, while minimizing effects on data loading randomness and model convergence.

    \item At the CP level, we implement a fine-grained per-document sharding method with an adaptive sharding selection mechanism to achieve the most efficient sharding for each input micro-batch at runtime.

    \item We conduct comprehensive evaluations and demonstrate that \textit{\Mname} achieves an average speedup of $1.23\times$ across various model scales and context window sizes.
\end{itemize}

%% file: sections/02_background.tex
\section{Background}
In this section, we provide the background of 4D parallelism LLM training and the cause of workload imbalance at different parallelism hierarchies. 

\begin{figure}[t] \small
    \centering
     \includegraphics[width=0.99\columnwidth]{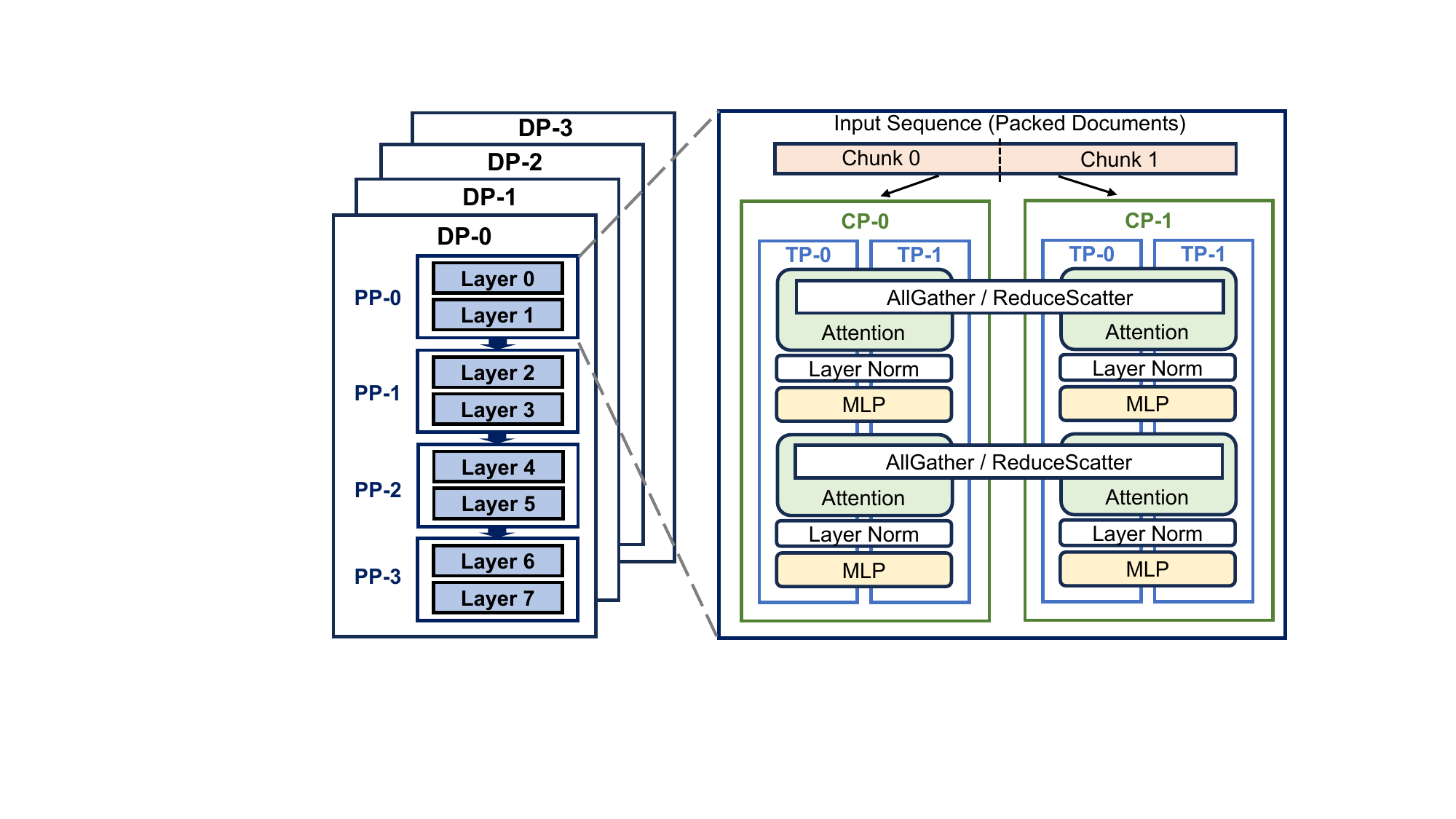}   
    \caption{Overview of 4D parallelism for LLM training.}
    \label{fig: 4d_parallel}
\end{figure}

\subsection{4D Parallelism LLM Training}
Training extremely large LLMs with billions or even trillions of parameters is challenging, involving a significant amount of engineering effort to tune the multi-level parallelism~\cite{zheng2022alpa}.
The state-of-the-art LLM training framework features a 4-dimensional parallelism~\cite{harper2019nemo, jacobs2023deepspeed}, including data parallelism (DP)~\cite{rajbhandari2020zero}, pipeline parallelism (PP)~\cite{narayanan2019pipedream}, context parallelism (CP)~\cite{nvidia2023megatron}, and tensor parallelism (TP)~\cite{shoeybi2019megatron}. Figure~\ref{fig: 4d_parallel} presents example of a (TP=2, CP=2, PP=4, DP=4) 4D parallelism. 

\textbf{Data Parallelism}: In DP, the input global batch is partitioned and distributed across DP workers, with each worker owning a part of the global batch. By default, model parameters are duplicated across DP workers. Some advanced techniques like ZeRO~\cite{rajbhandari2020zero} and FSDP~\cite{ott2021fully} partition model parameters, gradients, and optimizer states across DP workers to reduce memory consumption. During training, each DP worker computes parameter updates using its local batch independently and then synchronizes gradients with other workers via AllReduce (or ReduceScatter when using FSDP).

\textbf{Pipeline Parallelism}: Within each DP worker, the devices are further partitioned into multiple PP workers through pipeline parallelism. In PP, the model is split in a layer-wise manner, with each PP worker owning several chunks of layers. The input batches of a DP worker are also divided into multiple micro-batches. During training, a micro-batch traverses through all PP workers from first to last, and then reverses direction in the backward pass. Peer-to-peer (P2P) communication is required to send activations and gradients during forward and backward passes, respectively.

\textbf{Context Parallelism}: {CP is designed to address the large memory consumption of activations in long-context training. As shown in Figure~\ref{fig: 4d_parallel}, CP duplicates the model parameters but shards the input and activations along the sequence length dimension across CP workers. 
CP operates either in a ring-based manner using P2P communication~\cite{liu2023ring} or involves AllGather communication during the forward phase to collect the KV (key and value) tensors of all tokens, and ReduceScatter communication for the gradients of the KV tensors in backward~\cite{dubey2024llama}. All other operators, such as Linear and LayerNorm, operate independently on all workers, similar to DP.}

\textbf{Tensor Parallelism}: TP splits the model parameters within a layer (e.g., attention, FFN) and distributes them across multiple TP workers. TP is often applied in conjunction with Sequence Parallelism (SP)~\cite{korthikanti2023reducing} to further split the input tensor and activations. In this paper, when referring to TP, we mean both TP and SP by default. With TP, each GPU only has part of the input and parameters, resulting in intensive AllGather and ReduceScatter communication during training. Therefore, TP is typically applied within a single node, while other levels of parallelism are applied across nodes.

\begin{figure}[t] \small
    \centering
     \includegraphics[width=0.99\columnwidth]{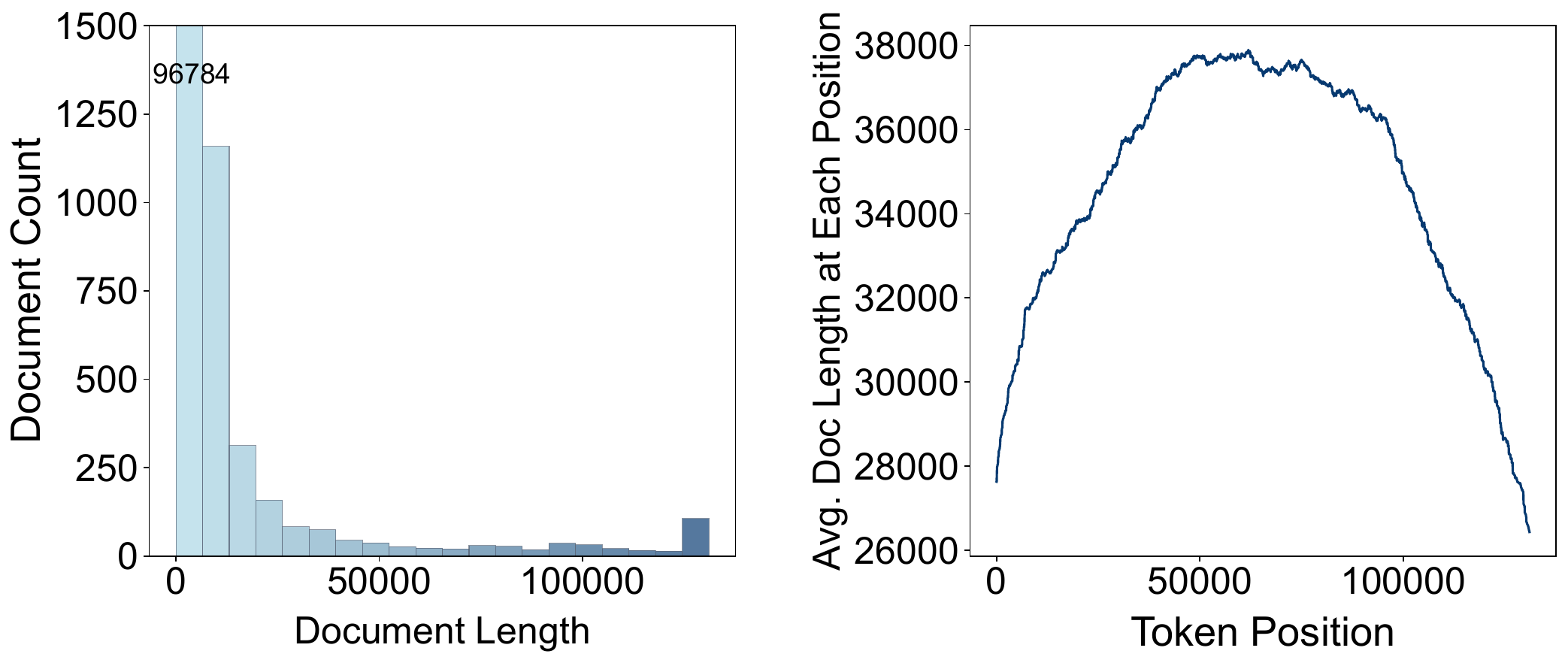}   
    \caption{Input Data Statistics: Distribution of input document lengths (Left) and average document length at each token position (Right).}
    \label{fig: input_distribution}
\end{figure}

\subsection{Varying Input Document Length}
In large-scale LLM training, the lengths of input documents could vary significantly, especially when using a large context window. 
To illustrate this, we profile the characteristics of training data in our 128K context length training job, as shown in Figure~\ref{fig: input_distribution}.
\textbf{From a per-document perspective}, input document lengths distribution is highly skewed. As shown in the left part of Figure~\ref{fig: input_distribution}, the majority of input documents are relatively short, while some extremely long documents exist, with the longest reaching the full context window size. The presence of an extremely long document in an input batch can easily lead to significant workload imbalances across micro-batches.
\textbf{From a per-token perspective}, we calculate the average document length corresponding to each position within the context window. As shown in Figure~\ref{fig: input_distribution} (Right), there is an approximately $1.4\times$ difference in average document length across token positions. 
Tokens in the middle are more likely to belong to longer documents. This is because when the data loader packs input documents into batches, long documents that happen to extend beyond the context window size are truncated at the tail to ensure the total sequence length matches the context window size.
This figure clearly illustrates the heterogeneity in per-token computation intensity during long-context LLM training.
These observations highlight the need for an input-aware solution that can dynamically balance workloads by accounting for variations in document length and per-token arithmatic intensity.

%% file: sections/03_motivation.tex
\section{Motivation}\label{sec: motivation}

{To motivate the design of \textit{\Mname}, we first conduct an in-depth analysis of workload imbalances occurring at different parallelism hierarchies in 4D parallelism LLM training.
We then introduce a baseline solution that shuffles and repacks input documents across batches. Lastly, we discuss why this baseline solution cannot fully eliminate workload imbalances and investigate the tradeoff between input packing balance and model convergence.}

\begin{figure}[t] \small
    \centering
    \subfloat[Imbalance Analysis (TP=8, CP=16, PP=16, DP=4): (1) Normalized computation latency (group by DP and PP); (2) Normalized computation latency in a CP group.]{
     \includegraphics[width=1\columnwidth]{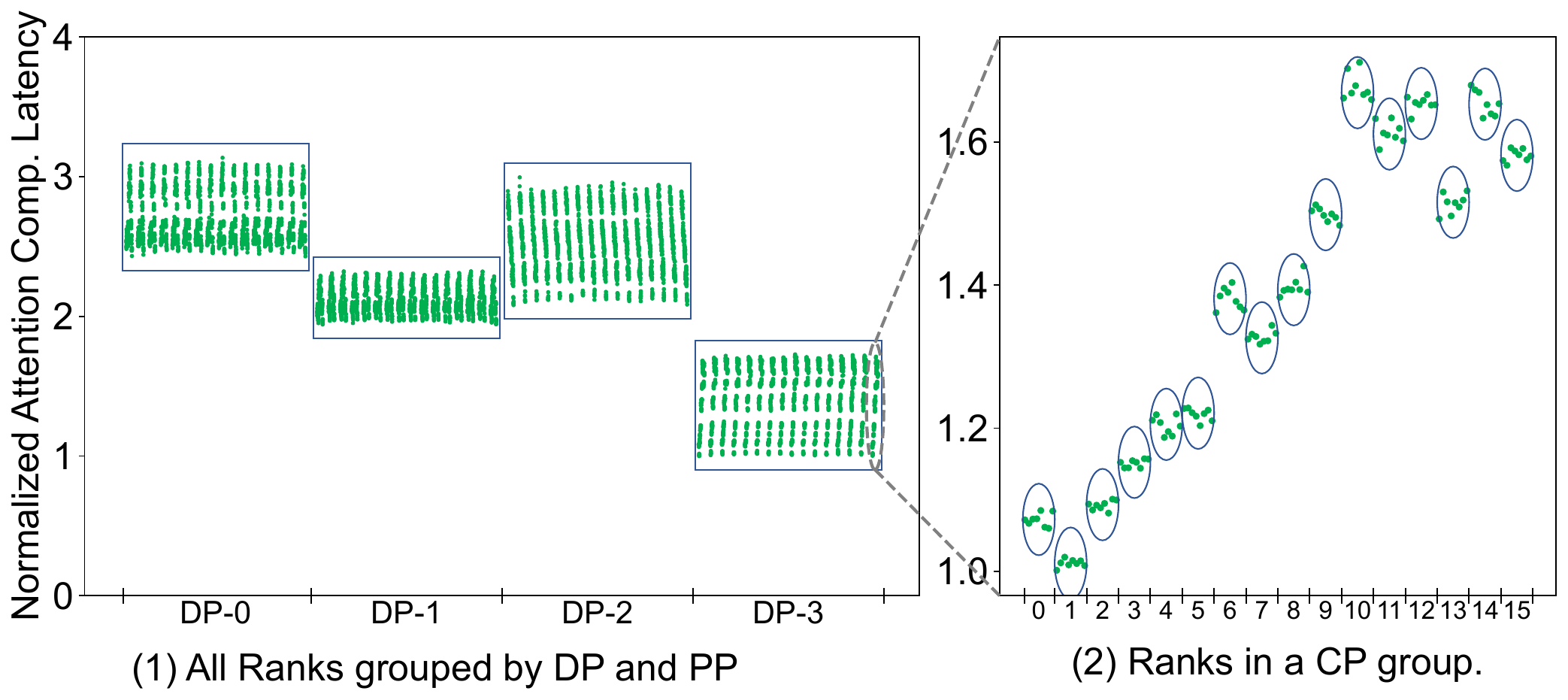}} 
     \vspace{0.1em}
     \subfloat[Document packing at PP level and sequence sharding at CP level.]{
     \includegraphics[width=1\columnwidth]{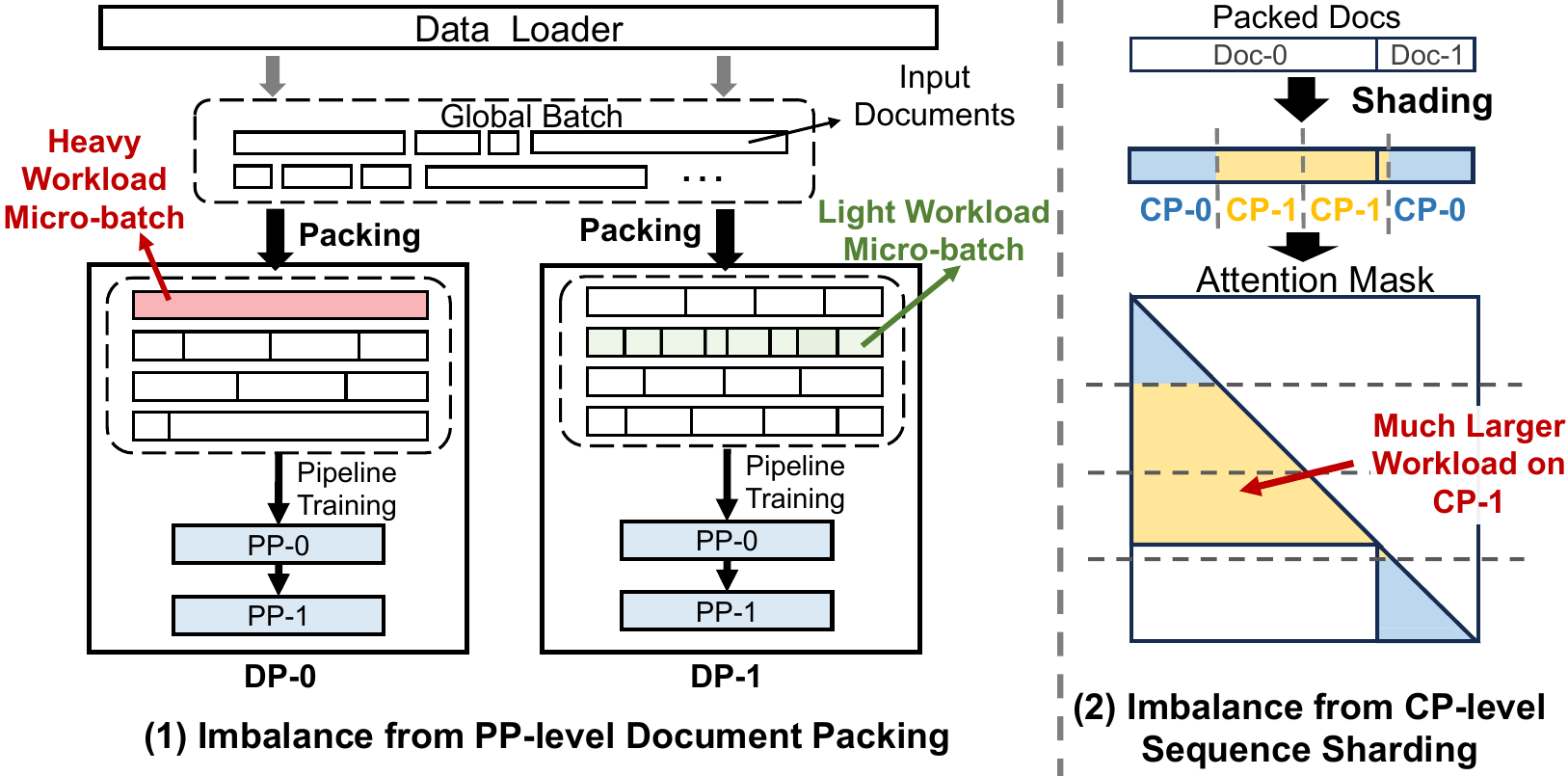}}
    \caption{The workload imbalance comes from the PP-level document packing and CP-level sequence sharding.}
    \label{fig: imba_analysis}
\end{figure}

\subsection{Imbalance Analysis}
After analyzing the performance trace from our internal 8K-GPU 128K context window size training job, we identify two primary causes of workload imbalance: (1) workload imbalance among micro-batches at the PP level, and (2) imbalance across sequence shards at the CP level. 
To demonstrate this, we accumulate the attention computation latency on each GPU. The results have been given in Figure~\ref{fig: imba_analysis}. 

\textbf{PP Level Imbalance:} 
{As shown in Figure~\ref{fig: imba_analysis}~(a)(1), the attention computation latency across DP workers varies significantly. Within each DP worker, we observe ``vertical lines'' formed by the data points, each representing a PP worker within a DP worker. Each PP worker shows nearly identical workloads, as all PP workers within a DP worker process the same set of micro-batches. 
Based on the results shown in Figure~\ref{fig: imba_analysis}~(a)(1), we conclude that the workload imbalance at both the DP and PP levels stems from the imbalance across micro-batches, which is caused by the input packing process. As illustrated in the left part of Figure~\ref{fig: imba_analysis}~(b), input documents are packed into sequences (micro-batches) of equal length. While this fixed-length packing strategy ensures that each PP and DP worker processes the same number of tokens, variations in per-token computation intensity result in workload imbalances at these levels. For instance, a micro-batch containing only a single long document (highlighted in red) has larger workload compared to a micro-batch composed of multiple shorter documents (highlighted in green).}

\begin{figure}[t] \small
    \centering
     \includegraphics[width=0.99\linewidth]{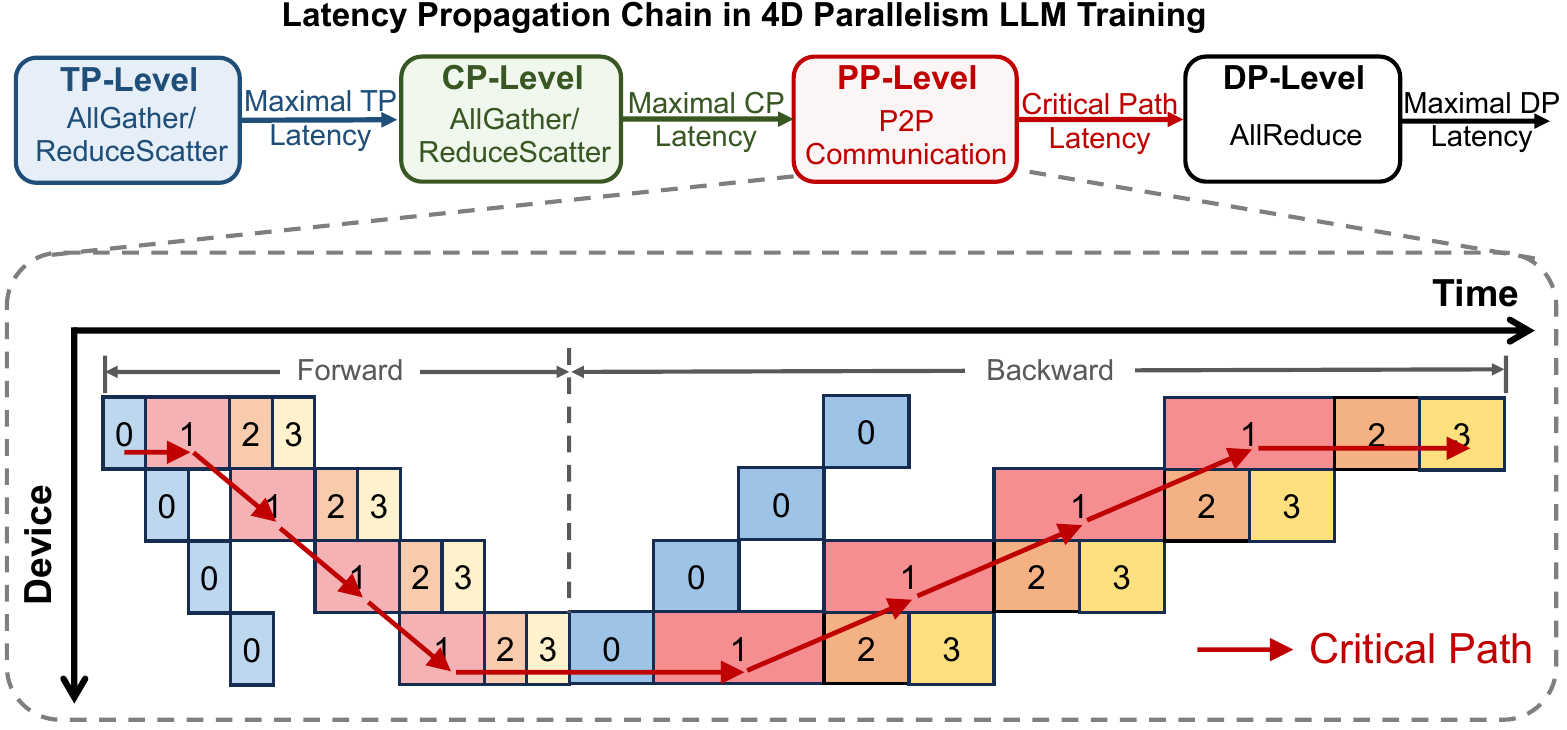}   
    \caption{The process of latency propagation in 4D parallelism LLM training across different parallelism hierarchies. The impact of workload imbalance is enlarged at the PP level.}
    \label{fig:propagation}
\end{figure}

\textbf{CP Level Imbalance:} 
{To better illustrate the imbalance at the CP level, we zoom in on a specific PP worker (also referred to as a CP group). As shown in Figure~\ref{fig: imba_analysis}~(a)(2), significant workload imbalances are observed across CP workers (indicated by circles), while the TP workers within each CP worker exhibit similar computation latencies (data points within each circle).
This imbalance across CP workers comes from the sequence sharding at the CP level. As shown in Figure~\ref{fig: imba_analysis}~(b)(2), an input sequence is divided into chunks with an equal number of tokens, which are then distributed to CP workers. 
The state-of-the-art approach partitions the input sequence into \( 2 \times CP\_size \) chunks. The \( i \)-th CP worker is assigned the \( i \)-th and \( (2 \times CP\_size - 1 - i) \)-th chunks to achieve better load balancing~\cite{dubey2024llama}. This sharding strategy works well when the sequence only has a single document. However, if the sequence is packed with multiple documents, it may lead to significant workload imbalances across CP workers, as demonstrated in the figure.  
Although sequence chunks are further divided and distributed among TP workers, no imbalance is observed at the TP level. This is because, prior to computation, all TP workers perform an AllGather operation to collect the entire sequence chunk. As a result, each TP worker within a CP worker processes the same sequence chunk, eliminating imbalances at the TP level.}

\textbf{Imbalance Propagation:}\label{sec: propogation}
{During training, the workload imbalance will be propagated from inner-level parallelism to outer-level parallelism. The imbalance will be accumulated and amplified, finally leading to a significant impact on end-to-end training latency. 
In DP, CP, and TP, collective communication~\cite{nccl} such as AllReduce, AllGather, and ReduceScatter is performed during training. All workers in these parallelism hierarchies work in a synchronized manner. As a result, the training latency of a given DP, CP, and TP group is determined by the slowest worker within that group.
In contrast, at the PP level, different PP workers serve as producers and consumers of each other. As shown in Figure~\ref{fig:propagation}, the critical path of PP is the latency of the largest micro-batch traversing all PP workers plus the forward and backward passes of remaining micro-batches on the first PP worker. The distinct data dependency relationship between PP workers amplifies the imbalance, resulting larger impact on total latency.
Due to the imbalance propagation, higher micro-batch training latency can arise from two main reasons: (1) Imbalances propagated from inner-level hierarchies (e.g., CP sharding imbalances); (2) Micro-batches with inherently larger workloads due to PP-level packing.
This highlights the importance of eliminating the imbalance in all parallelism hierarchies.} 

\begin{figure}[t] \small
    \centering
    \includegraphics[width=0.99\columnwidth]{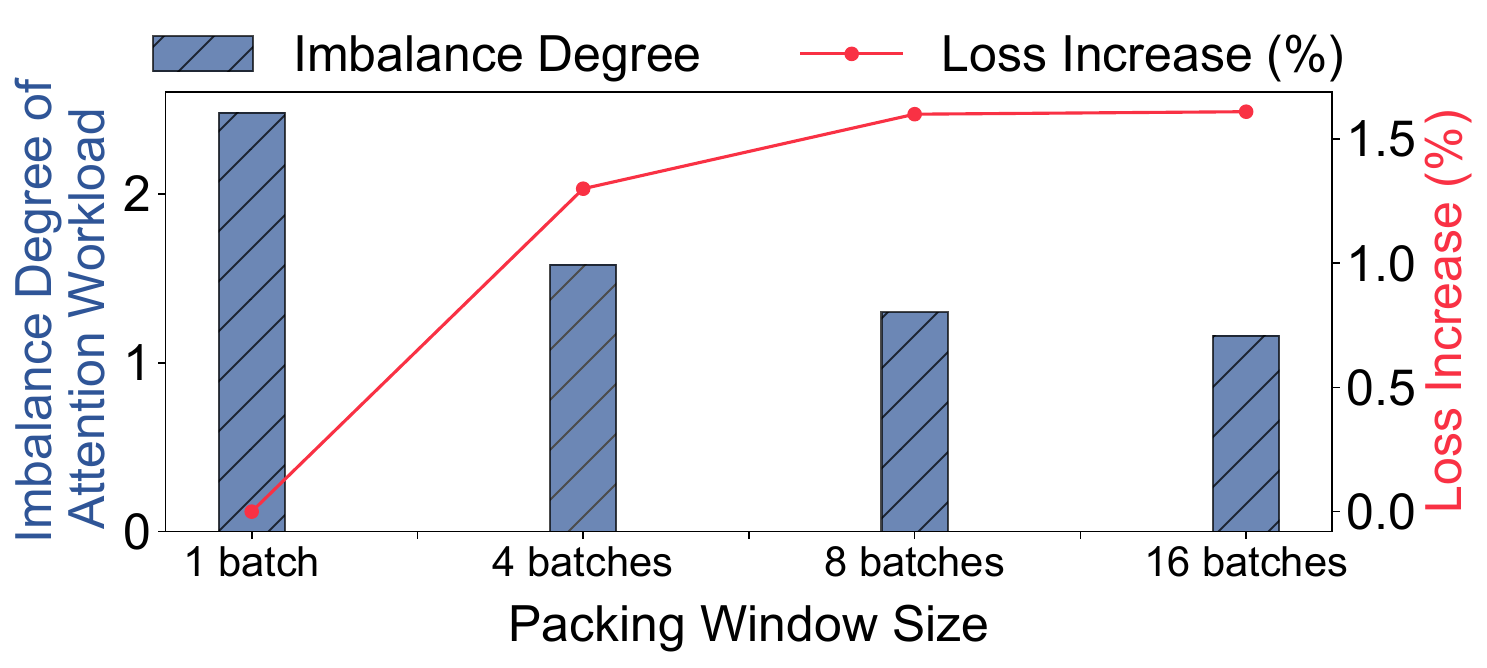}   
    \caption{A larger packing window improves workload balance but leads to an increase in training loss.}
    \label{fig: tradeoff}
\end{figure}

\subsection{Baseline Solution: Fixed-Length Packing}\label{sec: fix_len}
A potential solution to address the workload imbalance issue is to optimize the packing of input documents.
Current 4D parallelism frameworks require all micro-batches to have a uniform sequence length~\cite{dubey2024llama}, equal to the context window size, to enable efficient batching of input sequences.
Building on this fixed-length requirement, we implement a baseline shuffling and packing optimization that packs input documents into micro-batches of equal size.
Formally, given a set of input documents from one or more global batches, the objective is to pack these documents into several micro-batches with a fixed total length and balance the attention computation workload among all micro-batches.
Without loss of generality, we use the causal mask as an example to calculate attention workload.
With causal mask, the attention computation workload of a micro-batch is proportional to $\sum_{i=1}^N d_i^2$, where $d_i$ is the length of each document within the micro-batch.
The problem of searching the optimal document packing is NP-hard, as it can be extended from the classic Number Partition Problem~\cite{hartmanis1982computers} by adding a number sum constraint and a square-sum objective. 
To search for the optimal packing, we formulate the task as an integer linear programming (ILP) problem.
Assuming we have $N$ documents, each has length $d_i$, and we would like to pack these documents into $M$ micro-batches, each with a total length of $L$. The objective is to minimize the maximum workload among the micro-batches:

\begin{equation}\label{eq: fix-len}
\begin{array}{ll@{}ll}
\text{minimize}  & \max( \displaystyle\sum\limits_{i=1}^{N} x_{ij}\cdot d_i^2), &\ j=1 ,\cdots, M \\
\text{subject to} & \displaystyle\sum\limits_{j=1}^M  x_{ij} = 1, &\ i=1 ,\cdots, N\\
 & \displaystyle\sum\limits_{i=1}^N x_{ij}\cdot d_i \le L, &\ j=1 ,\cdots, M\\
 &  x_{ij} \in \{0,1\} &
\end{array}
\end{equation}
in which $x_{ij}$ is a binary variable representing the packing plan. Specifically, $x_{ij}=1$ means document $i$ is packed into micro-batch $j$. With this ILP formulation, we then use a commercial solver~\cite{gurobi} to obtain the optimal packing plan.

\subsection{Tradeoff Analysis}\label{sec: tradeoff}
Optimizing input document packing across more global batches could help achieve a higher degree of workload balance. 
However, it also manipulates the execution order of more input documents, affecting the randomness of data sampling and loading. This may negatively impact model quality and affect model convergence.
{To evaluate the tradeoff between packing balance and model quality, we pretrain a $550M$-parameter model for $52K$ steps using various packing window sizes.
We then assess the degree of workload imbalance of input batches after packing under different settings. The imbalance degree is calculated as \(\frac{Max\_Attn}{Avg\_Attn}\), where \(Max\_Attn\) represents the maximum attention computation workload in the global batch, and \(Avg\_Attn\) denotes the average attention computation workload of all micro-batches in the global batch.}
As shown in Figure~\ref{fig: tradeoff}, when optimizing the packing across a single global batch, the workload imbalance across micro-batches still remains high. 
If the number of global batches increases, the fixed-length packing optimization could achieve a better workload balance.
However, the final training loss increases as more global batches are involved in the packing optimization, due to reduced data loading randomness caused by repacking a larger number of input documents.
These results indicate that naïve fixed-length packing optimization cannot achieve a good workload balance without compromising model quality, highlighting the need for more sophisticated solutions.

{The tradeoff between packing balance and model quality motivates us to break the fixed-length constraint of micro-batches and design a more flexible packing strategy. In the following two sections, we will present the details of \textit{\Mname}, including the PP-level variance-length packing and heuristic outlier document delay optimization (\S\ref{sec: PP}) and the CP-level fine-grained and adaptive sharding optimization (\S\ref{sec: CP}).}

%% file: sections/04_PP_level.tex
\section{Var-Len Packing and Outlier Delay for PP}\label{sec: PP}

At the PP level, we focus on balancing the workload across micro-batches by repacking input documents in a workload-aware manner. \textbf{First}, we design a variable-length packing strategy to achieve a higher degree of workload balance within the same packing window size compared to fixed-length packing (\S\ref{sec: var-len packing}).
\textbf{Second}, we propose an outlier document delay method to adaptively delay the training of extremely long documents. This approach helps minimize the impact on data randomness while achieving near-optimal workload balance across micro-batches (\S\ref{sec: delay}).
\textbf{Finally}, we design and implement an efficient heuristic algorithm to optimize packing at runtime with negligible overhead (\S\ref{sec: packing algo}).

\begin{figure}[t] \small
    \centering
    \includegraphics[width=0.95\columnwidth]{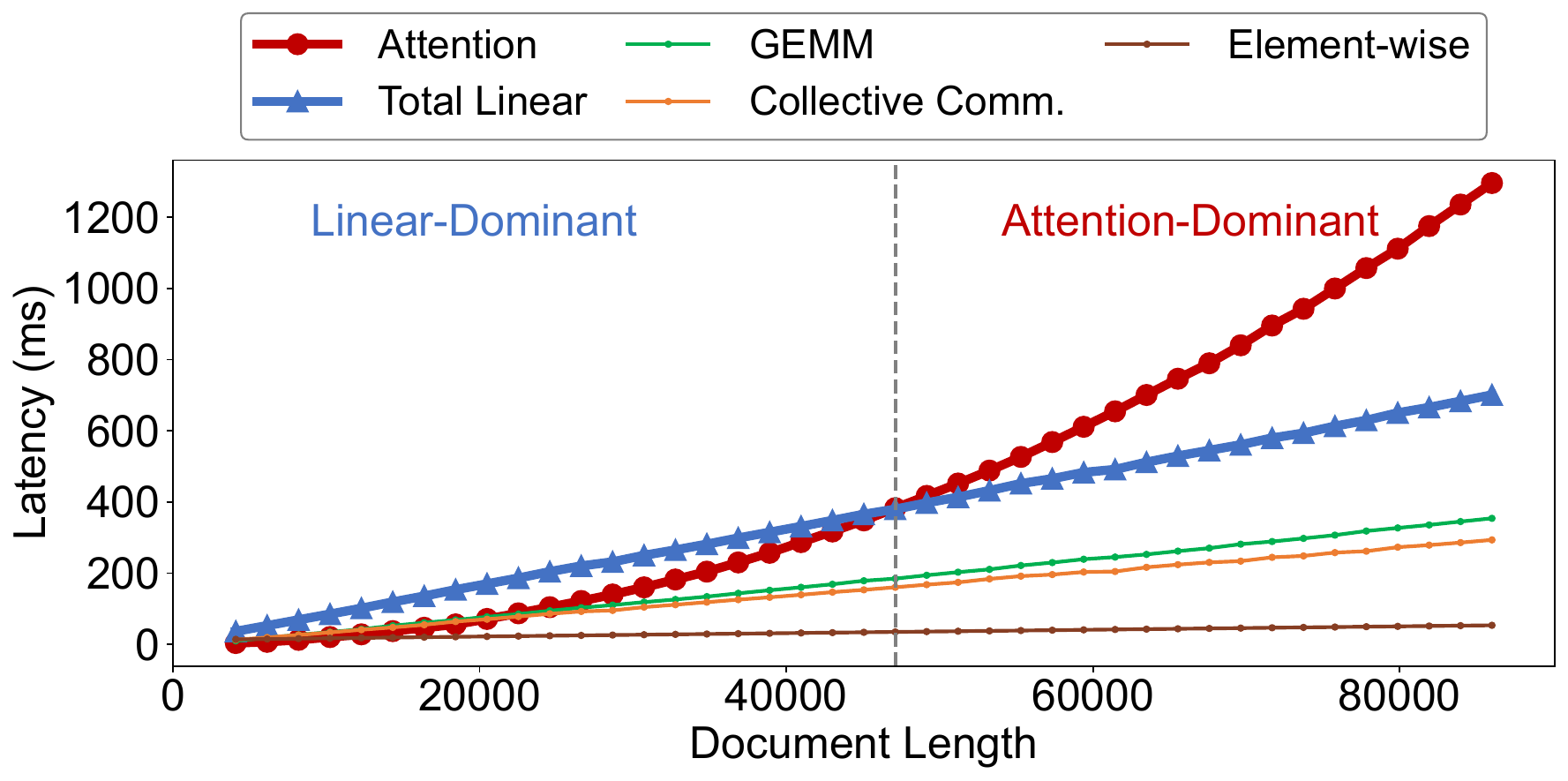}   
    \caption{The relationship between operation latency and the input document length. (Total Linear is the sum of GEMM, collective communication, and element-wise operators)}
    \label{fig: gemm_atten}
\end{figure}

\subsection{Workload-Aware Var-Length Packing}
\label{sec: var-len packing}
The main limitation of the baseline fixed-length packing is that it cannot achieve balance if there is an extremely long document in a global batch.
For example, if the length of a document equals the context window size, it becomes impossible to create another micro-batch consisting of shorter documents with an equal computation workload due to the quadratic complexity of attention computation.
To address this limitation, we propose a variable-length packing strategy, which allows each micro-batch to have a different sequence length. The key insight behind our design is that the workload of a micro-batch is not solely determined by attention computation. Other operations, such as GEMM computation, element-wise operations, and collective communication (e.g., AllGather and ReduceScatter), also contribute significantly to the training latency and are influenced by the documents within each micro-batch.
To demonstrate it, we present the relationship between operation latency and document length in Figure~\ref{fig: gemm_atten}.
It can be observed that attention computation latency increases quadratically with the length of input documents, while other operations, such as GEMM, collective communication, and element-wise operations, exhibit a linear relationship between operation latency and document length.

This relationship presents an opportunity to further improve workload balance beyond fixed-length packing. If a long document has significantly higher attention computation latency compared to other operations, we can pack multiple shorter documents together to extend the latency of other operations, thereby matching the total latency of the long document.
Specifically, we extend the fixed-length packing to a variable-length approach. The optimization goal shifts from balancing only the attention computation workload to balancing the total workload, including all operations:

\begin{equation}\label{eq: var-len}
\begin{array}{ll@{}ll}
\text{minimize}  & \max( \displaystyle\sum\limits_{i=1}^{N} (W_{a}(x_{ij}\cdot d_i) + W_{l}&(x_{ij}\cdot d_i))),  \\
& &\ j=1 ,\cdots, M  \\
\text{subject to} & \displaystyle\sum\limits_{j=1}^M  x_{ij} = 1, &\ i=1 ,\cdots, N\\
 & \displaystyle\sum\limits_{i=1}^N x_{ij}\cdot d_i \le L_{max}, &\ j=1 ,\cdots, M\\
 &  x_{ij} \in \{0,1\} &
\end{array}
\end{equation}
in which $W_a(\cdot)$ and $W_l(\cdot)$ are projection functions that compute the attention computation latency and the latency of all other operations, respectively, based on document length. Both $W_a(\cdot)$ and $W_l(\cdot)$ can be derived from offline profiling.
And $L_{max}$ represents the maximum sequence length permitted by GPU memory constraints.

\begin{figure}[t] \small
    \centering
    \includegraphics[width=0.99\columnwidth]{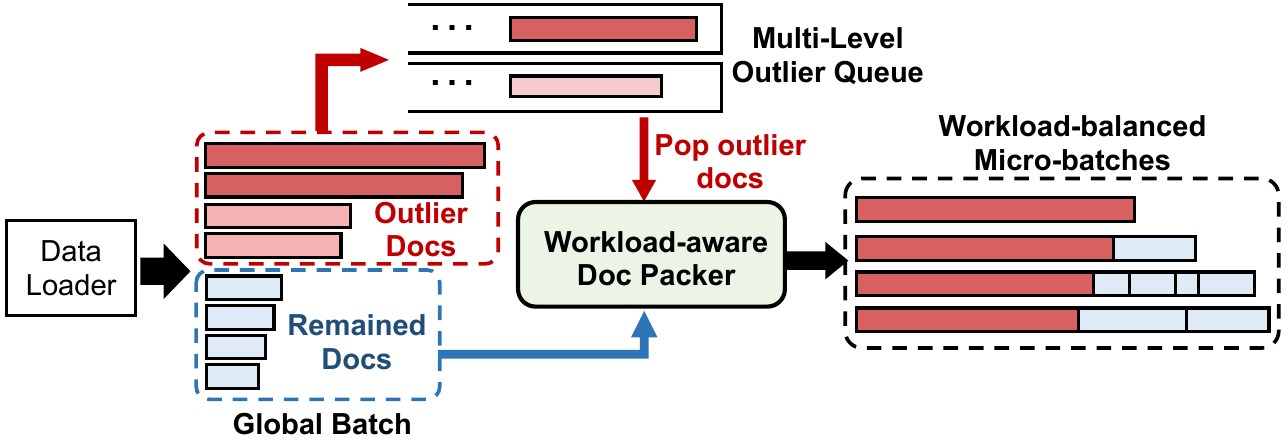}   
    \caption{The process of outlier document delay combined with var-length packing.}
    \label{fig: delay}
\end{figure}

\subsection{Outlier Document Delay}~\label{sec: delay}
Our var-length packing strategy enables a higher level of balance by allowing short documents to be packed into a longer sequence that exceeds the fixed context window size. However, within a single global batch, the number of short sequences may be insufficient to equalize the computation workload across all micro-batches. To address this challenge, we propose adaptively delaying the execution of outlier documents, i.e., extremely long documents. 
This approach is based on our observation that, although these outlier documents have a significant impact on workload imbalance, they only contribute a small proportion of tokens for training. By selectively delaying the training of tokens from a few extremely long documents, 
we could achieve a more balanced workload across micro-batches while minimizing the impact on data randomness.

\textbf{Design Details:} As illustrated in Figure~\ref{fig: delay}, we implement a multi-level waiting queue for outlier documents. Assuming we have $n$ outlier queues, each queue is associated with a hyperparameter $L_{i}$, which specifies the minimum length of documents in the $i$-th queue, where $L_1 < L_2 < \cdots < L_n$.
When a new global batch arrives, documents with lengths greater than $L_1$ are considered as outliers and are added to the corresponding waiting queue $i$, where $L_i \leq Doc\_len < L_{i+1}$. Here, $Doc\_len$ represents the length of the document.
The execution of documents in the outlier queue is delayed until the queue accumulates enough outlier documents. Once the queue size reaches the number of micro-batches, the outlier documents are popped from the queue and added to each micro-batch. This ensures workload balance across micro-batches, as each micro-batch contains exactly one outlier document.

\begin{algorithm}[t] \footnotesize
  \caption{Heuristic Var-length Packing Algorithm}
  \label{algo: packing}
\SetAlgoLined
  \SetKwInOut{Input}{input}
  \SetKwInOut{Output}{output}
  \Input{Dataloader: $\mathit{D}$, Waiting Queues: $\mathit{Q}$, \\
  \#Micro-Batch per Iteration: $N$, 
  \\ Sequence Length Upper bound: $L_{max}$}
  \Output{Packed Micro-Batches for Training: $\mathit{B}$}

    $\mathit{Remained\_Doc}$ = []; \\
   \For{$\mathit{Cur\_Batch}$ $\mathbf{in}$  $\mathit{D}$}{

     $\mathit{Doc\_Set}$ = $\mathit{Remained\_Doc}$;\label{line:doc_set_init} \\
     \For{$\mathit{Doc}$ $\mathbf{in}$  $\mathit{Cur\_Batch}$}{
        \tcc{Delay the execution of outlier documents.}
        
        \eIf{$\mathit{Doc}$.$\mathit{Is\_Outlier()}$}{
            $\mathit{Q}.{Add}(\mathit{Doc})$; \\
        }{
            $\mathit{Doc\_Set}.{Push}(\mathit{Doc})$; \\
        }
     }\label{line:loop_cur_batch_end}
    \For{$\mathit{q}$ $\mathbf{in}$ $\mathit{Q}$}{\label{line:loop_Q_start}
        \If{$\textbf{len}$($\mathit{q}$) $\ge$ $N$}{
          \tcc{Pop outlier documents for the current batch.}
          $\mathit{Doc\_Set}.{Push}(\mathit{q}.{Pop}(N))$;\\
        }
    }\label{line:loop_Q_end}

    \tcc{Sort the documents in descending order by length.}
    $\mathit{Doc\_Set}.Sort\_by\_Length()$;\label{line:sort_doc} \\ 
    \tcc{Start packing.}
    $\mathit{New\_Batch}$ = $\textbf{Create\_Batch}(N)$; \\
    \For{$\mathit{Doc}$ $\mathbf{in}$  $\mathit{Doc\_Set}$}{
     \tcc{Get micro-batches with minimum workload/length.}
     $\mathit{W\_idx}$ = $\mathit{New\_Batch}.Get\_Min\_Workload()$; \\
     $\mathit{L\_idx}$ = $\mathit{New\_Batch}.Get\_Min\_Length()$; \\

     \eIf{$\mathit{New\_Batch[W\_idx]}.Len()$ + $\mathit{Doc}.Len()$ < $L_{max}$}{
        $\mathit{New\_Batch[W\_idx]}.Push(\mathit{Doc})$; \\
     }{
        \eIf{$\mathit{New\_Batch[L\_idx]}.Len()$ + $\mathit{Doc}.Len()$ < $L_{max}$}{
            $\mathit{New\_Batch[L\_idx]}.Push(\mathit{Doc})$;\label{line:try_min_len} \\
        }{
          $\mathit{Remained\_Doc}.Push(\mathit{Doc})$;\label{line:remain_doc}\\
        }
     }
    }
    $\mathit{B}.Push(\mathit{New\_Batch})$;\\  
   }
\end{algorithm}

\subsection{Heuristic Packing Algorithm}~\label{sec: packing algo}
Although an ILP solver can derive the optimal packing for a given set of input documents, the solving time is impractically high. 
To address this, we design a heuristic algorithm that combines variable-length packing with outlier document delay optimization to efficiently produce packed micro-batches with balanced workloads.
As listed in Algorithm~\ref{algo: packing}, the algorithm takes a dataloader $\mathit{D}$, a set of waiting queues $\mathit{Q}$ for outlier documents working in a FIFO manner, the number of micro-batches $N$ per iteration, and an upper limit $L_{max}$ for sequence length as input. It will finally output a series of packed micro-batches with balanced workloads for each training iteration.
The packing process begins by iterating over the input batches in the dataloader $\mathit{D}$ and adding all outlier documents into the corresponding waiting queue in $\mathit{Q}$ for delayed processing (Line~\ref{line:doc_set_init}-\ref{line:loop_cur_batch_end}).
If any queue $\mathit{q}$ in $\mathit{Q}$ reaches size $N$, the documents in $\mathit{q}$ are popped out and added to the pending documents for the current batch $\mathit{Doc\_Set}$ (Line~\ref{line:loop_Q_start}-\ref{line:loop_Q_end}).
The packing operates in a greedy manner: pending documents are sorted in descending order by length. The algorithm first attempts to pack each document into the micro-batch with the minimal workload (calculated using \(W_a(\cdot)\) and \(W_l(\cdot)\) as defined in Equation~\ref{eq: var-len}), provided the total length remains within the upper limit $L_{max}$. 
If this is not possible, it then tries the micro-batch with the minimal length (Lines~\ref{line:sort_doc}-\ref{line:try_min_len}). Documents that cannot fit within these constraints are saved for the next iteration (Line~\ref{line:remain_doc}).
Our heuristic approach combines outlier document delay with variable-length document packing to efficiently balance computation workload across all micro-batches.

%% file: sections/05_CP_level.tex
\section{Fine-grained and Adaptive Sharding for CP} \label{sec: CP}

At the CP level, we aim to improve workload balance across document shards by implementing a fine-grained per-document sharding strategy, ensuring that each CP worker receives an equal computation workload (\S\ref{sec: per_doc}).
Additionally, we observe a tradeoff between attention kernel efficiency and sharding granularity (\S\ref{sec: cp kernel}). 
To maximize overall performance, we conduct an in-depth analysis and adaptively select the optimal sharding strategy for a given input batch (\S\ref{sec: cp_select}).

\begin{figure}[t] \small
    \centering
    \includegraphics[width=0.99\columnwidth]{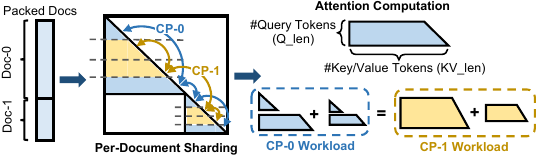}   
    \caption{Overview of fine-grained per-document sharding.}
    \label{fig: per-doc}
    \vspace{-10pt}
\end{figure}

\begin{figure}
    \centering
    \includegraphics[width=0.99\linewidth]{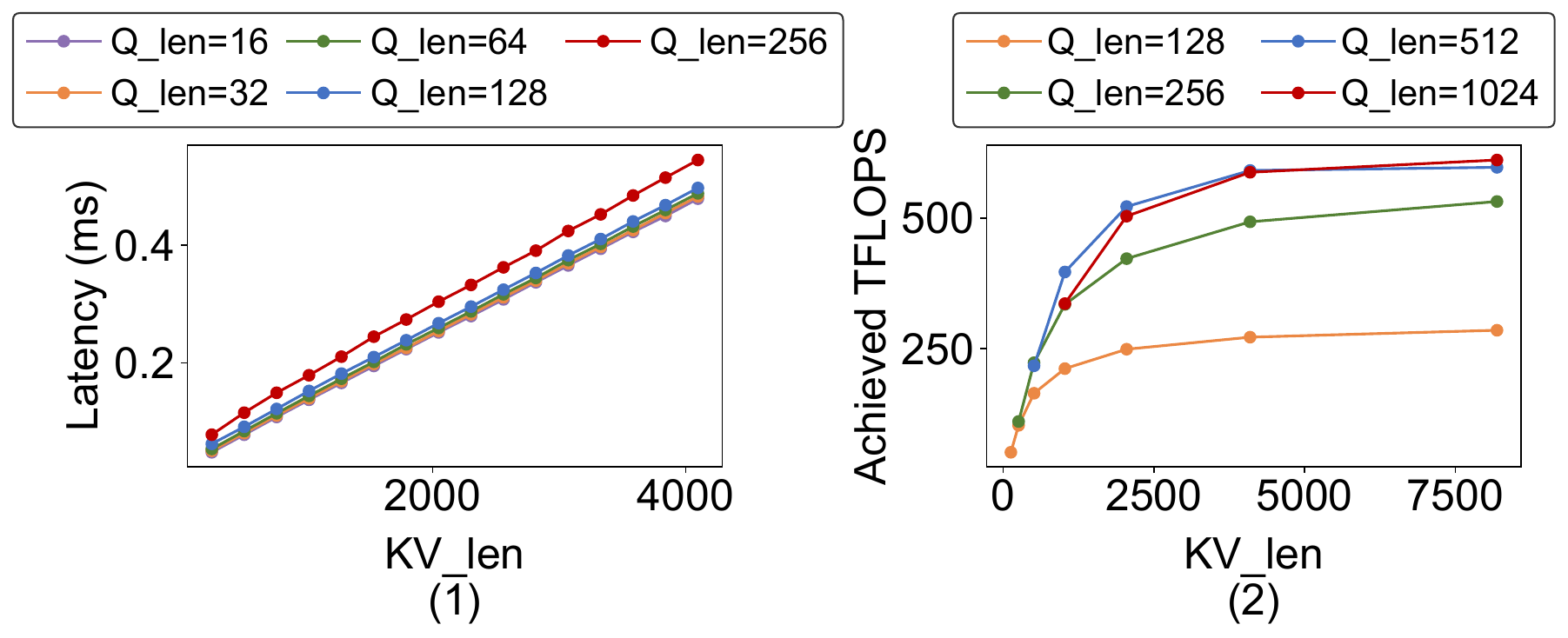}
    \caption{Attention kernel performance profiling: (Left) Attention forward latency; (Right) Achieved TFLOPs of the attention forward kernel.}
    \label{fig: atten_kernel}
    \vspace{-10pt}
\end{figure}

\subsection{Per-Document Sharding Design}\label{sec: per_doc}

At the CP level, the sequence of micro-batches is sharded across CP workers. Each CP worker works on an exclusive sequence shard. 
Existing CP implementation employs a \textbf{Per-Sequence Sharding} strategy, which equally shards the entire input sequence into $2\times CP\_size$ chunks. This method could easily lead to significant attention computation workload imbalance when the input sequence is packed with multiple documents.
To eliminate the workload imbalance issue at the CP level, we propose sharding the sequence in a fine-grained manner. Specifically, we conduct \textbf{Per-Document Sharding} to divide each document into $2\times CP\_size$ document chunks.
As shown in Figure~\ref{fig: per-doc}, each CP worker takes a symmetrical pair of document chunks for each input document. With our fine-grained per-document sharding strategy, each CP worker not only receives the same number of tokens (ensuring workload balance in GEMM computation and collective communication) but also gets the same attention computation workload.

\textbf{Avoid Padding}: 
Our fine-grained per-document sharding strategy divides each input document into \(2 \times CP\_size\) document chunks. However, document lengths are not always divisible by \(2 \times CP\_size\), requiring extending the document length through adding some padding tokens. To avoid the redundant computation introduced by document padding, we design a padding-free per-document sharding method.
Specifically, we split each document into two parts: one divisible by $2 \times CP\_size$ and the remaining tokens. Assuming the length of the $i$-th document is $L_i$, where $L_i = D_i + R_i$, with $D_i = \lfloor \frac{L_i}{2 \times CP\_size} \rfloor$. 
We apply the standard per-document sharding on the $D_i$ part, while the tokens in $R_i$ are distributed to CP workers in a round-robin fashion.
Since both $\sum_{i=1}^n L_i$ and $\sum_{i=1}^n D_i$ are both divisible by $2 \times CP\_size$, it follows that $\sum_{i=1}^n R_i = \sum_{i=1}^n (L_i - D_i)$ is also divisible by $2 \times CP\_size$. This ensures that each CP worker receives an equal number of tokens, thereby eliminating the need for padding.

\subsection{Kernel Efficiency vs. Sharding Balance} \label{sec: cp kernel}
Our per-document sharding strategy fully eliminates workload imbalance at the CP level. However, splitting each document into multiple shorter chunks may compromise kernel efficiency.
The kernel efficiency may drop with fine-grained sharding due to two major reasons: \textbf{(1) Tile-level Computation Wasting:} The computation of attention is split into smaller tiles and distributed to different thread blocks for execution on GPU. For example, in the attention forward kernel of FlashAttention~\cite{dao2022flashattention}, the tile size is set to 128. If the number of tokens is less than the tile size, the thread block will still perform the full computation on 128 tokens, which will waste a significant amount of computation. 
To illustrate this, we profile the attention forward latency for query token lengths ranging from 16 to 256. As shown in Figure~\ref{fig: atten_kernel} (Left), when the number of query tokens ($Q\_len$) increases from 16 to 128, the kernel latency remains constant. This is because all short documents with fewer than 128 tokens are padded to 128 tokens for computation at the kernel level. In contrast, as $Q\_len$ increases from 128 to 256, the kernel latency rises significantly.
\textbf{(2) Inefficient Tensor Memory Accelerator (TMA) Usage:} TMA is a feature introduced in the NVIDIA Hopper architecture that enables asynchronous memory copying between global memory and shared memory on GPUs~\cite{h100gpu}. With large document lengths (e.g., $Q\_len \ge 256$), multiple thread blocks process different Q tokens while sharing the same KV tokens of the document chunk. This allows KV tensor loading to be shared via the L2 cache using TMA load multicast, significantly reducing the latency of transferring KV tensors from global memory to shared memory.
As shown in Figure~\ref{fig: atten_kernel} (Right), when the length of the Query tensor increases from 128 to 256, the achieved TFLOPs increase significantly, demonstrating the impact of leverage TMA load multicast. 

\begin{figure}[t]
    \centering
    \includegraphics[width=0.99\linewidth]{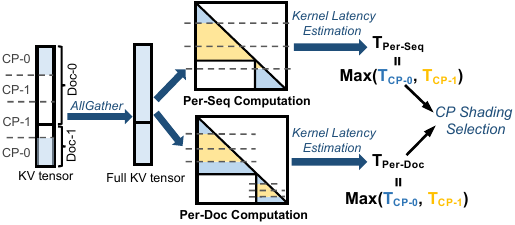}
    \caption{The process of adaptive CP sharding selection.}
    \label{fig: CP_adpative}
    \vspace{-5pt}
\end{figure}

These profiling results clearly demonstrate the tradeoff between attention kernel efficiency and CP sharding balance. If per-document sharding is applied to input sequences composed of short documents, it can introduce redundant computation at the kernel level and reduce the achieved TFLOPs, ultimately leading to longer attention computation latency.

\subsection{Adaptive Sharding Selection}\label{sec: cp_select}

Based on our analysis in Section~\ref{sec: cp kernel}, although fine-grained per-document sharding achieves optimal workload balance at the CP level, it does not necessarily guarantee better performance, as the attention kernel may become less efficient with the more fine-grained document chunks generated by per-document sharding. 
To address this issue, we propose to adaptively select the optimal sharding strategy for each micro-batch at runtime. As shown in Figure~\ref{fig: CP_adpative}, during the forward pass at the CP level, an AllGather communication is performed across CP workers to collect the full KV tensor. We then calculate the input tensor shapes for the attention kernel (number of query tokens and key/value tokens) in the per-sequence and the per-document sharding cases. Finally, we predict the attention kernel latency and select the CP sharding strategy that yields lower attention computation latency. 

To accurately estimate the attention kernel latency, we leverage the insights provided in \S\ref{sec: cp kernel}. 
First, we calculate the total floating point operations required for the attention computation. The kernel-level padding is also considered by padding the document chunk to a multiple of the tiling size. 
Next, we estimate the achieved TFLOPs for the given tensor shape using the data collected from offline profiling which includes the impact of TMA usage. Finally, the attention kernel latency is estimated by dividing the amount of floating point operations by the achieved TFLOPs.
By adaptively selecting CP sharding, \textit{\Mname} minimizes the CP level training latency.

%% file: sections/06_evaluation.tex
\section{Evaluation}

In this section, we evaluate \textit{\Mname} across various LLM sizes and 4D parallelism configurations, with model sizes ranging from \textit{550M} to \textit{70B}. We begin by presenting the improvements in end-to-end training latency. Next, we analyze the speedup contributions from individual optimizations at the PP level (\S\ref{sec: PP}) and the CP level (\S\ref{sec: CP}), respectively. Finally, we demonstrate that the optimizations of \textit{\Mname} do not compromise model quality or model convergence by comparing the training loss curve.

\begin{figure*}[t]
    \centering
    \includegraphics[width=0.99\linewidth]{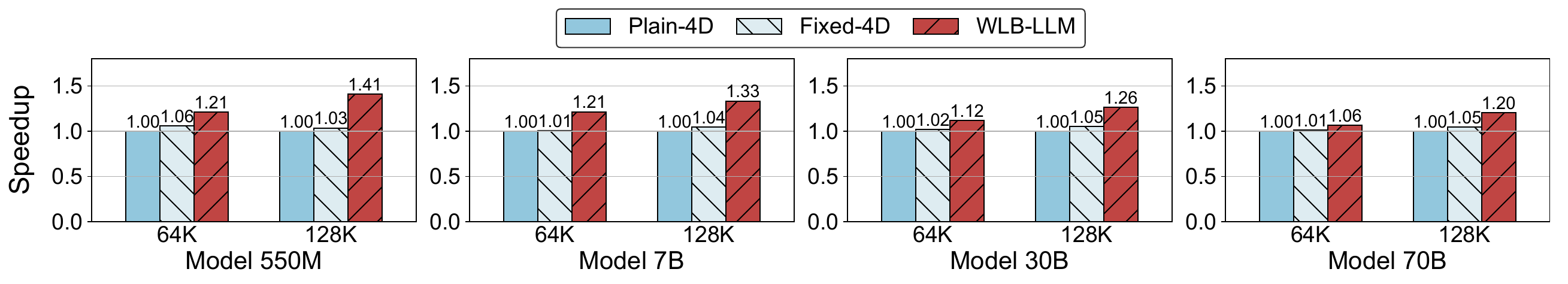}
    \caption{Training performance speedups of \textit{\Mname{}} and \textit{Fixed-4D} over \textit{Plain-4D} across various configurations.}
    \label{fig:speedup}
    \vspace{-5pt}
\end{figure*}

\begin{table}[t]
\centering
\scalebox{0.8}{
\begin{tabular}{cccc}
\Xhline{3\arrayrulewidth}
\textbf{\makecell{Model Size}} & \textbf{\makecell{Context\\Window}} & \textbf{\makecell{\#GPU}} & \textbf{\makecell{4D Parallelism Configs\\(TP, CP, PP, DP)}} \\  \Xhline{3\arrayrulewidth}
\multirow{2}{*}{\textit{550M}} & \textit{64K}  & \textit{32}  & \textit{(2, 2, 4, 2)} \\ 
                      & \textit{128K} & \textit{32}  & \textit{(2, 4, 4, 1)} \\ \Xhline{2\arrayrulewidth}
\multirow{2}{*}{\textit{7B}}   & \textit{64K}  & \textit{32}  & \textit{(4, 2, 4, 1)} \\ 
                      & 128K & \textit{64}  & \textit{(8, 2, 4, 1)} \\ \Xhline{2\arrayrulewidth}
\multirow{2}{*}{\textit{30B}}  & \textit{64K}  & \textit{64}  & \textit{(8, 2, 4, 1)} \\ 
                      & 128K & \textit{128} & \textit{(8, 4, 4, 1)} \\ \Xhline{2\arrayrulewidth}
\multirow{2}{*}{\textit{70B}}  & \textit{64K}  & \textit{256} & \textit{(16, 4, 4, 1)} \\ 
                      & \textit{128K} & \textit{256} & \textit{(16, 4, 4, 1)} \\ \Xhline{3\arrayrulewidth}
\end{tabular}
}
\caption{Model and 4D parallelism configurations.}
\label{tab:4D_parallelism}
\end{table}

\subsection{Experiments Setup}
\textbf{Hardware:} We deploy \textit{\Mname} on a cluster with 32 nodes. Each node is equipped with $8\times$ NVIDIA H100 SXM 80GB GPUs interconnected via high-bandwidth NVLink, while cross-node communication is facilitated by RDMA over Converged Ethernet (RoCE).

\vspace{5pt}
\noindent \textbf{Models and Parallelism Configurations:} We conduct experiments on a series of our internal LLaMA-like models, spanning four different scales: $550M$, $7B$, $30B$, and $70B$. The $7B$ model shares the same architecture as the LLaMA2-7B model~\cite{touvron2023llama}. The other models retain the same architecture while proportionally adjusting the number of layers and model dimension size.
For each model, we evaluated performance using two different context window sizes: $64K$ and $128K$. Each model scale and context window size is associated with a corresponding 4D parallelism configuration. 
The global batch size is set to $PP\_size \times DP\_size$ and we use \textit{bfloat16} precision for all evaluations.
Details of the training setup and parallelism configurations are provided in Table~\ref{tab:4D_parallelism}.
When mapping 4D parallelism to the hardware, inner-level parallelism dimensions (e.g., TP or CP) are prioritized for mapping to intra-node GPUs, leveraging the high-bandwidth NVLink for efficient communication. Outer-level parallelism dimensions, such as DP, are subsequently mapped across multiple nodes.
Throughout the rest of the paper, we use \textit{Model Size-Context Window Size} to denote a specific configuration. For instance, \textit{7B-128K} refers to the $7B$ model with a $128K$ context window.

\vspace{5pt}
\noindent \textbf{Baselines:} We compare \textit{\Mname} with two baselines: 
\begin{itemize}
    \item \textbf{Plain-4D}: This is our internal codebase for large-scale LLM training, supporting 4D parallelism to enable efficient training and scaling up to $100K$ GPUs. \textit{Plain-4D} directly uses input batches obtained from the dataloader for training without optimizing the input packing. For CP sharding, \textit{Plain-4D} employs a per-sequence sharding method, which shards the input sequence at the whole-sequence level.
    \item \textbf{Fixed-4D}: \textit{Fixed-4D} applies the baseline fixed-length packing optimization, as described in Section~\ref{sec: fix_len}. To minimize packing overhead, a greedy algorithm is used instead of the solver, and the packing size is restricted to a single global batch to preserve data loading randomness and prevent an increase in training loss. For CP sharding, \textit{Fixed-4D} utilizes a fixed sharding strategy throughout the entire training process, either per-sequence or per-document. We then use the better-performing result for comparison.
\end{itemize}

\subsection{Training Performance}
We run \textit{\Mname} and all baselines on diverse model size and context window size: 

\vspace{3pt}
\noindent \textbf{Plain-4D vs. Fixed-4D}: As shown in Figure~\ref{fig:speedup}, \textit{Fixed-4D} achieves only marginal improvements over \textit{Plain-4D}, with an average speedup of approximately $1.03\times$ across all settings. This limited gain is primarily because \textit{Fixed-4D} adjusts document packing only within a single global batch and is constrained by the context window size. It fails to address the presence of outlier documents with extremely long lengths (e.g., a document with a length equal to the context window size). Given that attention computation scales quadratically with document length, packing multiple short documents within the constraints of the context window cannot match the computation workload of an extremely long document.
Moreover, \textit{Fixed-4D} employs either per-sequence or per-document sharding at the CP level for all input batches. This approach overlooks the tradeoff between attention kernel efficiency and sharding balance, resulting in suboptimal performance. These limitations constrain the improvements of \textit{Fixed-4D} over the \textit{Plain-4D} baseline.

\begin{figure}[t]
    \centering
    \includegraphics[width=0.95\linewidth]{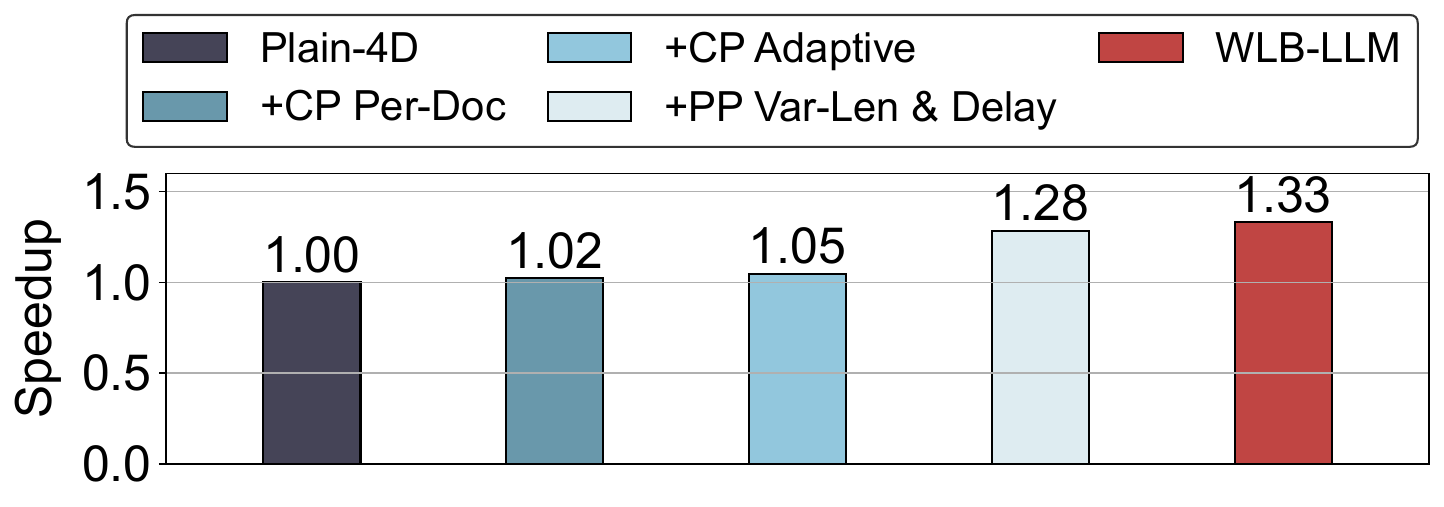}
    \caption{Performance breakdown of \textit{\Mname{}} on the 7B model with a 128K context window.}
    \vspace{-10pt}
    \label{fig:breakdown}
\end{figure}

\noindent \textbf{{\Mname} vs. Baselines}: 
Figure~\ref{fig:speedup} shows that \textit{\Mname} consistently outperforms all baselines across various model sizes and 4D parallelism configurations. Specifically, \textit{\Mname} achieves speedups of $1.23\times$ and $1.19\times$ over \textit{Plain-4D} and \textit{Fixed-4D}, respectively.
The significant improvement stems from two key optimizations in \textit{\Mname}. At the PP level, \textit{\Mname} employs a heuristic variable-length document packing algorithm that is not constrained by the fixed context window size. This algorithm also selectively delays the training of outlier documents, achieving a higher degree of workload balance compared to fixed-length packing. At the CP level, \textit{\Mname} combines per-sequence and per-document sharding, adaptively selecting the optimal sharding strategy for each micro-batch to maximize the performance. 

\noindent \textbf{Across Model Size and Context Window Size}: 
As shown in Figure~\ref{fig:speedup}, \textit{\Mname} achieves slightly lower speedup on larger models. This is because larger models involve more GPUs for training, which increases the ratio of communication latency to computation latency, making the impact of workload imbalance in the attention layer less significant.
On the other hand, increasing the context window size from 64K to 128K improves the average speedup from $1.15\times$ to $1.30\times$, as longer contexts exacerbate workload imbalance issue. A more detailed sensitivity analysis on the context window size is provided in Section~\ref{subsec:eval:opt_analysis}.

\begin{figure}[t]
    \centering
    \includegraphics[width=0.95\linewidth]{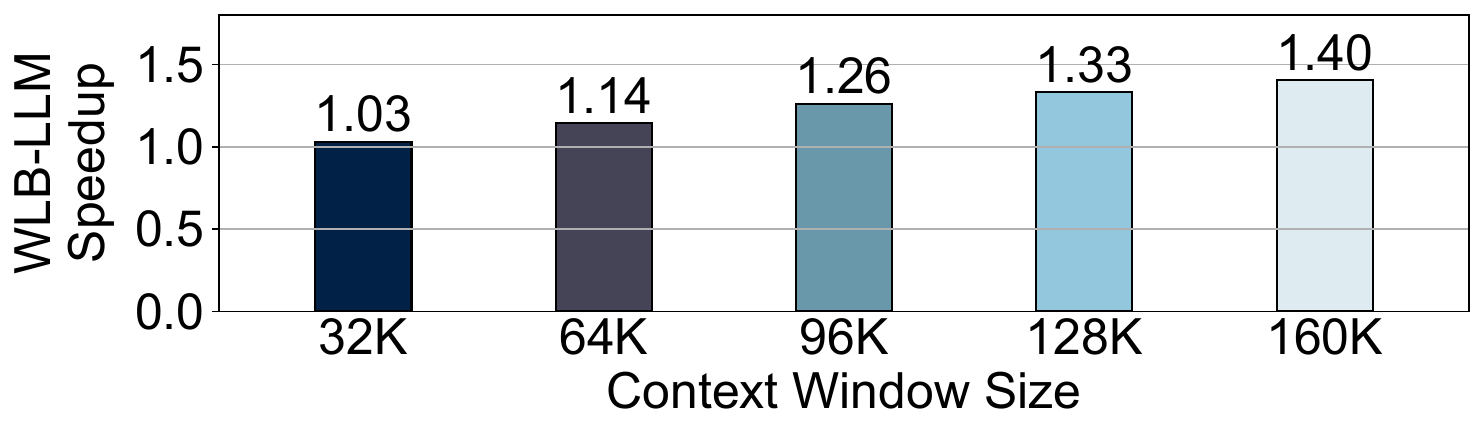}
    \caption{Speedups of \textit{\Mname{}} on the 7B model across context window sizes.}
    \label{fig:sensitive}
    \vspace{-10pt}
\end{figure}

\subsection{Optimization Analysis}
\label{subsec:eval:opt_analysis}

In this section, we conduct the optimization analysis of \textit{\Mname{}}. Specifically, we first report the performance breakdown to evaluate the impact of each optimization technique.
Second, we perform a sensitive study by measuring the speedups with different context window sizes.

\noindent \textbf{Speedup Breakdown:}
We show the speedup breakdown in Figure~\ref{fig:breakdown} by separately applying each optimization technique proposed in \textit{\Mname{}} to \textit{Plain-4D} for the \textit{7B-128K} training job.
By utilizing the fine-grained per-document sharding at the CP level, we observe a $1.02\times$ speedup due to the reduced imbalance among CP workers.
The speedup is limited due to the trade-off between the kernel efficiency and the sharding balance. As a result, performing per-document sharding is not always beneficial and can potentially cause performance degradation.
By adaptively selecting the per-document and per-sequence sharding strategies, we can improve the speedup to $1.05\times$.
We then apply the PP-level optimizations on \textit{Plain-4D} to study their effect. It can be observed that combining the heuristic var-length packing with the outlier documents delay offers a significant speedup of $1.28\times$.
Finally, we incorporate both CP and PP-level optimizations to maximally reduce the workload imbalance across the all parallelism hierarchies, leading to a final speedup of $1.33\times$.

\noindent \textbf{Speedup across Context Window Sizes:}
We investigate the impact of context window size on the performance improvements delivered by \textit{\Mname{}}.
Figure~\ref{fig:sensitive} depicts the speedups over \textit{Plain-4D} on $7B$ model across different context window sizes, varying from $32K$ to $160K$.
We observe that as the context window size grows, the achieved speedup becomes more significant, reaching $1.40\times$ with a 160K context window.
This is because a larger context window raises the likelihood of outlier documents appearing. Additionally, a larger context window also increases the proportion of attention computation and exacerbates the impact of imbalance on training latency.
The trend of increasing speedups demonstrates the significant potential of \textit{\Mname{}} in handling the ever-expanding context window sizes today.

\begin{table}[t]
\centering
\scalebox{0.8}{
\begin{tabular}{c|c|c|c}
\Xhline{3\arrayrulewidth}

\multicolumn{2}{c|}{\textbf{Packing Method}}           & \multirow{2}{*}{\textbf{\makecell{Imbalance\\Degree}}} & \multirow{2}{*}{\textbf{\makecell{Packing\\Overhead (ms)\\}}} \\ \Xcline{1-2}{1pt}
\multicolumn{1}{c|}{\textbf{Method}} & \textbf{Config} &                                                &                     \\ \Xhline{3\arrayrulewidth} 

\textit{Original Packing} & / & {1.44} & {0} \\ \Xhline{2\arrayrulewidth} 

\multirow{4}{*}{\textit{Fixed-Len Greedy}} & {\#global batch=1} & {1.41} & {4} \\ \cline{2-4} 
                                       & {\#global batch=2} & {1.22} & {5} \\ \cline{2-4} 
                                        & {\#global batch=4} & {1.11} & {5} \\ \cline{2-4} 
                                         & {\#global batch=8} & {1.08} & {5} \\ \Xhline{2\arrayrulewidth} 

\multirow{3}{*}{\textit{Fixed-Len Solver}} & {\#global batch=1} & {1.40} & {467} \\ \cline{2-4} 
                                         & {\#global batch=2} & {1.18} & {1488} \\ \cline{2-4} 
                                         & {\#global batch=4} & {1.09} & {25313} \\ \Xhline{2\arrayrulewidth} 

\multirow{3}{*}{\textit{WLB-LLM}} & {\#queue=1} & {1.24} & {8} \\ \cline{2-4} 
                                         & {\#queue=2} & {1.05} & {20} \\ \cline{2-4} 
                                         & {\#queue=3} & {1.05} & {23} \\ \Xhline{3\arrayrulewidth} 
\end{tabular}
}
\caption{Packing imbalance degree and overhead analysis.}
\label{tab: packing}
\end{table}

\subsection{Ablation Studies}
In this section, we first analyze the effectiveness of the packing and sharding optimization in \textit{\Mname}. Furthermore, we demonstrate that the system optimizations in \textit{\Mname} do not compromise model quality or slow down convergence.

\noindent \textbf{Packing Balance and Overhead Analysis:} 
To assess the balance degree of computation workload across micro-batches with different packing strategies, we profile and compare the forward latency of each micro-batch in a \textit{7B-128K} training job with different packing methods and configurations applied. 
As discussed in Section~\ref{sec: propogation}, the PP level latency is primarily determined by the largest micro-batch. Therefore, we use the following metric to represent the imbalance degree of a given batch: $\frac{Max\_Latency \times PP\_size}{Total\_Latency}$, where $Max\_Latency$ is the forward latency of the largest micro-batch and $Total\_Latency$ is the total forward latency of all micro-batches. A lower imbalance degree indicates that the given batch is more balanced in terms of workload. The result of the imbalance degree and the packing overhead (per-batch packing latency) of different methods have been given in Table~\ref{tab: packing}.
We evaluate four different packing methods: (1) \textit{Original Packing}, which uses the original input batch loaded from the dataloader; (2) \textit{Fixed-Len Greedy} is the packing method used in \textit{Fixed-4D} baseline, which shuffles documents in several global batches in a greedy manner to optimize workload balance across micro-batches; (3) \textit{Fixed-Len Solver}, which employs an ILP solver~\cite{gurobi} to solve Equation~\ref{eq: fix-len} and provides the optimal packing based on the given global batches; and (4) \textit{\Mname}, which utilizes var-len packing combined with outlier document delay. Additionally, we demonstrate the performance of \textit{\Mname} with different numbers of outlier document queues.

As shown in Table~\ref{tab: packing}, \textit{Fixed-Len Greedy} can slightly mitigate workload imbalance when packing across a single global batch. Packing across multiple global batches helps to achieve lower imbalance degrees, while it will incur higher training losses, as demonstrated in Figure~\ref{fig: tradeoff}.
As for \textit{Fixed-Len Solver}, it achieves lower imbalance degrees compared to \textit{Fixed-Len Greedy} under the same number of global batches. However, the solver-based solution suffers from significant packing overhead. For instance, when packing across 4 global batches, the average packing latency for each batch exceeds 25 seconds.
In contrast, \textit{\Mname} is the only solution which achieves both near-optimal imbalance degree and low packing overhead. 
For example, when having two outlier queues, \textit{\Mname} achieves $1.05$ imbalance degree. Additionally, the per-batch packing latency is only $20$ ms which is less than $0.65\%$ when compared to the per-step training latency.

\begin{figure}[t]
    \centering
    \includegraphics[width=0.99\linewidth]{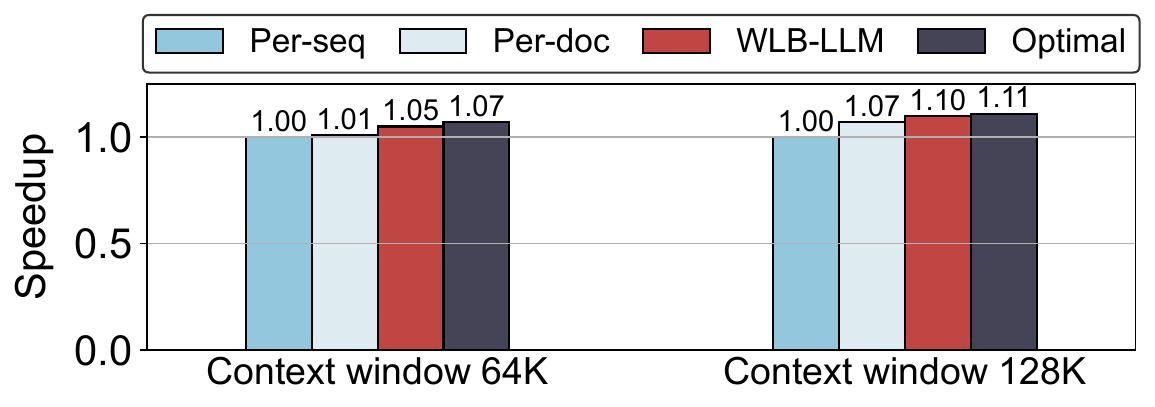}
    \caption{CP sharding performance comparison.}
    \vspace{-10pt}
    \label{fig: cp_sharding}
\end{figure}

\vspace{5pt}
\noindent \textbf{CP Sharding Performance Analysis:} 
To demonstrate the effectiveness of the per-document sharding and adaptive sharding selection optimization at CP level in \textit{\Mname}, we conduct a case on a single transformer layer in \textit{7B} model with CP size of 4. We compare the forward and backward latency with different sharding strategies including: (1) \textit{Per-Sequence Sharding (Per-Seq)}, (2) \textit{Per-Document Sharding (Per-Doc)}, (3) \textit{\Mname}, which determines the sharding between \textit{Per-Seq} and \textit{Per-Doc} adaptively based on the given micro-batch at runtime, and (4) \textit{Optimal}, which is the optimal result. It always chooses the sharding from \textit{Per-Seq} and \textit{Per-Doc} that yields lower latency.

As shown in Figure~\ref{fig: cp_sharding}, 
\textit{Per-Document Sharding} achieves a speedup of $1.01\times$ and $1.07\times$ over the \textit{Per-Sequence Sharding} baseline under context window sizes of 64K and 128K, respectively.
These results demonstrate the effectiveness of 
\textit{Per-Document Sharding} in reducing the workload imbalance at the CP level. 
However, the fine-grained \textit{Per-Document Sharding} may sacrifice kernel efficiency, particularly when the input sequence consists of many short documents.
To overcome this issue, \textit{\Mname} leverages an adaptive sharding selection to intelligently choose a better sharding strategy for each micro-batch at runtime. 
The results in Figure~\ref{fig: cp_sharding} show that \textit{\Mname} achieve $7.5\%$ and $3.4\%$ improvement over static \textit{Per-Sequence} or \textit{Per-Document} sharding.
Furthermore, \textit{\Mname} is very close to the optimal result, demonstrating the effectiveness of our adaptive selection method.

\begin{figure}[t]
    \centering
    \includegraphics[width=0.95\linewidth]{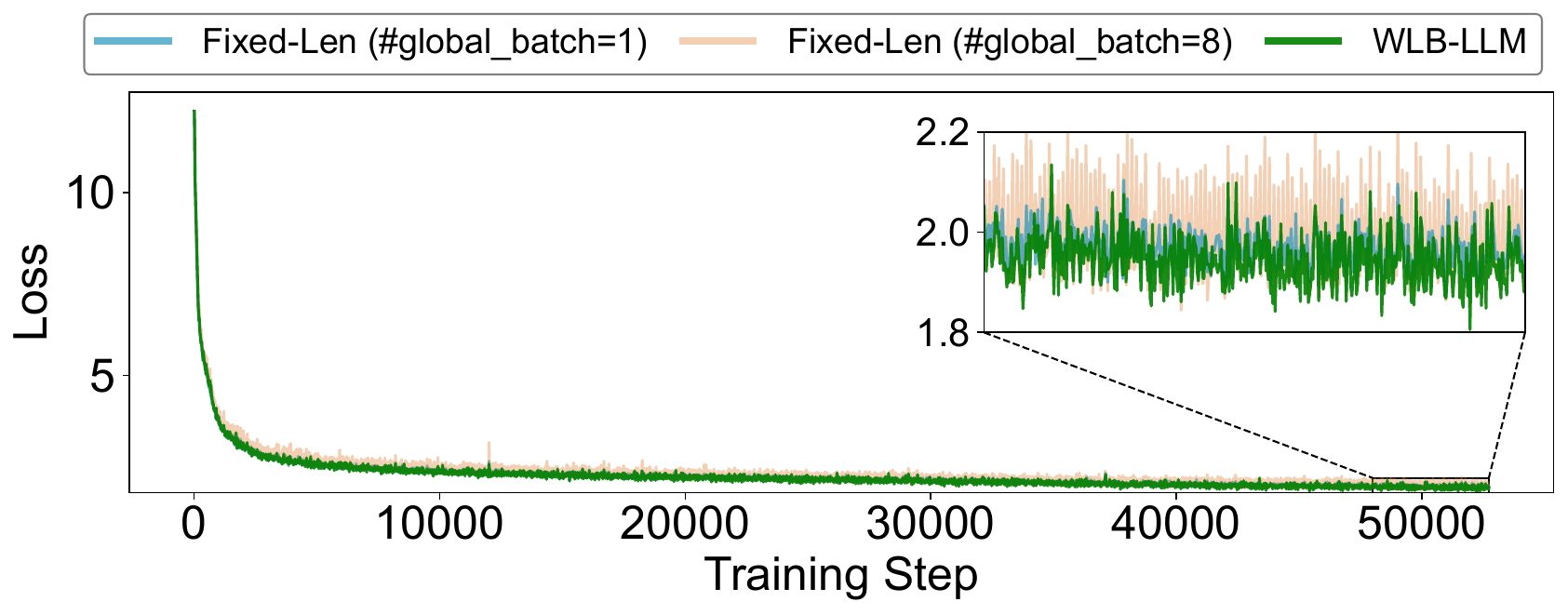}
    \caption{Training loss comparison on a \textit{550M} model.}
    \label{fig: loss}
    \vspace{-5pt}
\end{figure}

\vspace{5pt}
\noindent \textbf{Model Convergence Analysis:}
\textit{\Mname} employs a heuristic variable-length packing optimization that adjusts document packing and delays the execution of outlier documents. To demonstrate that this optimization does not affect model convergence, we present the training loss curve of our \textit{550M} model. 
As shown in Figure~\ref{fig: loss}, packing across 8 global batches results in a noticeable increase in training loss ($1.6\%$ on average). This is because packing over multiple global batches 
disrupts the randomness of data sampling in the dataloader, which leads to a different data distributed per batch than desired.
In contrast, \textit{\Mname} follows almost the same trend as fixed-length packing across a single global batch, since the heuristic variable-length packing algorithm used by \textit{\Mname} only delays outlier documents, which contributes to a small proportion of all input tokens. According to our profiling, each token is delayed by an average of $0.5$ iterations in \textit{\Mname}. This minimal delay preservers original data distribution at best, allowing \textit{\Mname} to improve training efficiency without compromising model quality.

%% file: sections/07_related_work.tex
\section{Related Work}

\textbf{Distributed LLM Training Frameworks:}
To address the challenges of training extra-large LLMs, prior approaches primarily rely on 3D parallelism~\cite{shoeybi2019megatron, narayanan2021efficient, team2020deepspeed, korthikanti2023reducing, wang2023zero++, zheng2022alpa, unger2022unity}, which integrates tensor parallelism~\cite{krizhevsky2012imagenet, shoeybi2019megatron, korthikanti2023reducing}, pipeline parallelism~\cite{huang2019gpipe, li2021terapipe, narayanan2021efficient, lamy2023breadth}, and data parallelism~\cite{rajbhandari2020zero, ren2021zero, zhao2023pytorch}.

Although 3D parallelism has proven its effectiveness in scaling model size, enabling the training of models with trillions of parameters~\cite{team2020deepspeed}, it struggles to scale context window size effectively.
To address this limitation, a new dimension of parallelism—context parallelism—has been introduced~\cite{nvidia2023megatron}, forming the 4D parallelism training paradigm.
Context parallelism splits the input context into chunks, effectively reducing the memory bottleneck associated with very long sequence-length inputs. 
Initially, context parallelism employs a ring-based approach that overlaps communication and computation using P2P communication~\cite{liu2023ring}. More recent approaches leverage collective communication methods (e.g., AllGather or AlltoAll) to aggregate key and value tensors, offering greater flexibility and better support for various types of attention masks~\cite{dubey2024llama, jacobs2023deepspeed, nvidia2023megatron}.
The introduction of context parallelism enables efficient training of LLMs with long context windows.
However, existing 4D parallelism frameworks overlook the heterogeneity in per-token arithmetic intensity, leading to significant workload imbalances across GPUs.

\vspace{5pt}
\noindent \textbf{Input Padding and Packing for LLM Training:}
The input of LLMs consists of samples with varying lengths. To batch input documents together and optimize GPU utilization during LLM training, the input documents must be organized into tensors of identical lengths. This can be achieved through two primary approaches: \textit{Padding}~\cite{shoeybi2019megatron} and \textit{Packing}~\cite{krell2021efficient}.
Padding involves adding zero padding to shorter documents within a micro-batch. However, this approach inevitably introduces redundant computation, communication, and memory overhead. To address this issue, prior work has focused on designing more efficient kernels to reduce redundant computation~\cite{zhai2023bytetransformer, zeng2022boosting} or optimizing batching strategies to minimize padding~\cite{jiang2024dynapipe}.
To fully eliminate redundant computation, recent works propose packing short documents together to form a single long input sequence~\cite{krell2021efficient}. After packing, an additional attention mask must be applied to ensure tokens only attend to others within the same document~\cite{flexattention, dubey2024llama}.
Due to its efficiency in avoiding redundant computation, packing has become the mainstream choice in LLM training. For example, LLaMA-3 adopts input packing~\cite{dubey2024llama} in its training process, and state-of-the-art high-performance attention implementations (e.g., FlashAttention~\cite{dao2022flashattention, dao2023flashattention}) also support efficient attention computation with document packing.
\textit{\Mname} focuses on input packing and addresses the workload imbalance issue that arises when input documents are packed together.

%% file: sections/08_conclusion.tex
\section{Conclusion}

In this paper, we presented \textit{\Mname}, a workload-balanced 4D parallelism framework for LLM training.
\textit{\Mname} identifies and addresses workload imbalances across different parallelism hierarchies.
At the pipeline parallelism level, \textit{\Mname} introduces a novel heuristic variable-length document packing algorithm that effectively mitigates workload imbalance across micro-batches. At the context parallelism level, \textit{\Mname} proposes a fine-grained and adaptive per-document sharding strategy to achieve the optimal training performance with sequence sharding.
Comprehensive experiments demonstrate that \textit{\Mname} outperforms existing 4D parallelism frameworks across various model sizes and parallelism configurations, achieving an average speedup of $1.23\times$.